# Wideband Spectrum Acquisition for UAV Swarm Using the Sparse Coding Fourier Transform

Kaili Jiang, Kailun Tian, Hancong Feng, Junyu Yuan, and Bin Tang

*Abstract*—As the trend towards small, safe, smart, speedy and swarm development grows, unmanned aerial vehicles (UAVs) are becoming increasingly popular for a wide range of applications. In this letter, the challenge of wideband spectrum acquisition for the UAV swarms is studied by proposing a processing method that features lower power consumption, higher compression rates, and a lower signal-to-noise ratio. Our system is equipped with multiple UAVs, each with a different sub-sampling rate. That allows for frequency backetization and estimation based on sparse Fourier transform theory. Unlike other techniques, the collisions and iterations caused by non-sparsity environments are considered. We introduce sparse coding Fourier transform to address these issues. The key is to code the entire spectrum and decode it through spectrum correlation in the code. Simulation results show that our proposed method performs well in acquiring both narrowband and wideband signals simultaneously, compared to the other methods.

## I. INTRODUCTION

UNMANNED aerial vehicle (UAV) has rapidly developed in recent decades and is gaining popularity in both military and civilian applications [1], such as surveillance, emergency response, communication, etc. By utilizing the small, low cost, and large numbers, the UAV swarm can significantly expand the environment sensing and mission completion capabilities through information sharing within the swarm [2]-[3].

Manuscript received April 28, 2023; revised month date, year; accepted month date, year. Date of publication month date, year; date of current version month date, year. This work was supported in part by the Key Project of the National Defense Science and Technology Foundation Strengthening Plan 173 under Grand 2022-JCJQ-ZD-010-12. The associate editor coordinating the review of this manuscript and approving it for publication was xxx. (*Corresponding author: Kaili Jiang*).

Kaili Jiang, Kaililun Tian, Hancong Feng, Junyu Yuan and Bin Tang are with the School of Information and Communication Engineering, University of Electronic Science and Technology of China, Chengdu, Sichuan, 611731, China (e-mail: jiangkelly@uestc.edu.cn; kailun_tian@163.com; 2927282941 @qq.com; Jyyuan@uestc.edu.cn; bint@uestc.edu. cn).

While UAV swarm technology is very promising, many outstanding issues need to be resolved. One such issue is the wideband spectrum sensing problem considered in this letter, and the main challenge is the high sampling rate, which results in huge data storage, transmission, and processing [4]. To monitor a wideband spectrum using a sweeping receiver [5], there is a long scanning time and a low probability of signal interception. Channelized receiver [6] has high power, serious crosstalk, and requires a large number of analog-to-digital converters (ADCs). Meanwhile, multi-coset sampling [7] and multi-rate sampling [8] utilize the parallel mode of multi-channel splicing, while the main difficulty lies in the high precision clock control for multi-chip ADCs [9]. Fortunately, the analog-to-information receiver based on the compressive sensing (CS) theory has been used to address this challenge, including random modulation [10], Nyquist folding receiver (NYFR) [11], modulated wideband modulator (MWC) [12], etc. However, the practical implementation of the existing CS-based spectrum sensing schemes is constrained by the need for high-speed pseudo-random sequences or pulse trains [13]. In addition, spectrum reconstruction has a strictly sparse restriction, high computational complexity, and high input signal-to-ratio (SNR) [14] requirements.

The existing schemes mentioned above are not suitable solutions for the UAV swarm, from the perspective of system structure, data processing, and transmission volume. Due to the exciting capability of sparse Fourier transform (SFT) [15]-[16], a novel wideband spectrum acquisition technique is proposed that combines UAV swarm and SFT. The implementation cost is significantly reduced due to sub-Nyquist sampling directly by multiple low-speed ADCs. Based on the system scenario and the SFT theory, coding the entire spectrum and decoding it through a spectrum correlation method in the code are proposed with non-strictly sparsity restriction. Unlike existing methods, the position perturbation caused by the UAVs is considered, and introduces the spectrum correlation instead of covariance sensing in the time domain. Simulation results are presented and compared with the existing techniques to demonstrate the superiority of the proposed scheme.

The remainder of this letter is organized as follows. The system scenario is given in section II. Then, the wideband spectrum acquisition method is proposed in section III. The simulation results are provided in section IV, and section V concludes the letter.

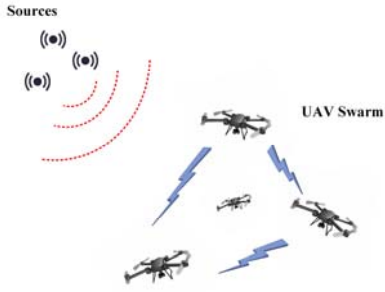

Fig. 1. Scenario of wideband spectrum acquisition using UAV swarm.

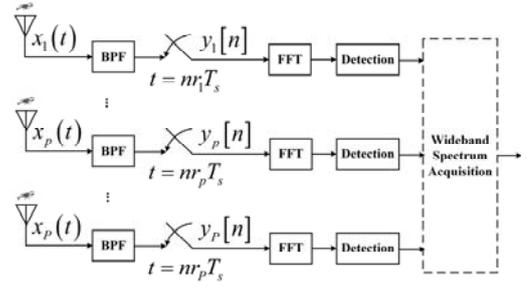

Fig. 1. Block diagram of the proposed joint spectrum acquisition system.

## II. SYSTEM SCENARIO

Consider a wideband spectrum acquisition scenario that uses $P$ UAVs, where the UAV swarm can be distributed in a non-uniform array. The schematic of the UAV swarm for wideband spectrum acquisition is shown in Fig. 1. Assume that there are $I$ uncorrelated and far-field signals over a wide frequency band colliding into the UAV swarm. Moreover, the process is wide-sense stationary with zero mean. Let $x_p(t), t \in \mathbb{R}$ combine the $I$ signals from the $p$ th UAV in the time domain, which can be expressed as

$$x_p(t) = \sum_{i=1}^{I} s_i(t - \tau_{i,p}) + n_p(t) \quad (1)$$

where $s_i(t)$ and $\tau_{i,p}$ denote the radio-frequency signal of interest and its time delay in the $p$ th UAV of the $i$ th source, respectively. And the term $n_p(t)$ denotes the additive white Gaussian noise (AWGN).

The far-field propagation model only considers the time delay in the UAV swarm. Meanwhile, each source signal has an unknown azimuth $\theta_i \in [-\pi/3, \pi/3]$ corresponding to the time delay. And the signals have distinct carrier frequencies $f_{c_i}$ confined within the bandwidth $B_s$ at one instant, which allows them to be distinguished from one another. Furthermore, the AWGN is an independent and identically distributed random variable, with zero mean and variance $\sigma^2$.

The output samples of the UAV swarm are obtained by $P$ uniform samplers with different sampling rates, and there is only one sampler for each UAV receiver. The block diagram of the proposed joint spectrum acquisition system is shown in Fig. 2. The inputs are digitized directly by the conventional ADC after passing through the preselected wide band pass filter (BPF) for the entire frequency range (2-18GHz). The set of swarm sampling intervals is $\{r_1 T_s, r_2 T_s, \ldots, r_P T_s\}$ determined by the Nyquist sampling rate $f_s = 1/T_s$. From the data acquisition perspective, the discrete-time random process $x_p[n], n \in \mathbb{N}$ of the $p$ th path can be unambiguously acquired by the Nyquist sampling rate $f_s \geq 2B_s$.

Without loss of generality, assuming $r_1 > r_2 > \ldots > r_P$. Then, the highest uniform sub-sampling rate of the system is $1/(r_P T_s) = f_s / r_P$. And the received outputs with the $P$ samplers can be respectively written as

$$\begin{aligned} y_1[n] &= x[nr_1] = x(nr_1 T_s) \\ y_2[n] &= x[nr_2] = x(nr_2 T_s) \\ &\vdots \\ y_P[n] &= x[nr_P] = x(nr_P T_s) \end{aligned} \quad (2)$$

Wideband spectrum sensing has been attempted under sparse sampling in numerous earlier papers. The focus of many other works is to estimate the frequency under a single path. The frequency of interest is then obtained using multiple sub-Nyquist samplers that are synchronized or precisely controlled by time delay. Then, by exploiting the temporal correlation between different samplers, the power spectrum can be reconstructed reliably in a low SNR environment without any sparsity requirement.

Compared to the single path scheme, array antennas can significantly improve the reliability of spectrum sensing by utilizing the spatial diversity gain. Unlike most others, the temporal correlation from different antennas cannot be utilized due to their non-synchronous nature. Therefore, a new method of wideband spectrum acquisition is needed under the UAV swarm scenario.

## III. WIDEBAND SPECTRUM ACQUISITION

The wideband spectrum acquisition of the signals received from the low-cost system will be realized by the sparse coding Fourier transform which will be introduced in the following.

### A. Frequency Bucketization

The first step starts by hashing the frequencies of the wideband spectrum into buckets. Following the basic property of the sampling theorem, subsampling in the time domain causes aliasing in the frequency domain. Thus, the low-speed subsampling from each UAV is a form of bucketization. Meanwhile, the frequencies hash to the same bucket through the aliasing for each UAV. Thus, the bucketization frequency for each interested $f_{c_i}$ can be denoted as

$$f_{l_p} = f_{c_i} - k_{p_i} f_{s_p} \quad (3)$$

within the range $(-f_{s_p}/2, f_{s_p}/2]$ for $p = 1, 2, \ldots, P$ with the Nyquist zone index $k_{p_i} \in \mathbb{Z}$ and the sampling rate $f_{s_p} = f_s / r_p$. Further, the value in each bucket is the sum of the aliasing values for the entire Nyquist zones. The focus is then on the occupied buckets and ignores the empty buckets.

*B. Frequency Estimation*

Next, the frequency and energy of the occupied buckets need to be identified for each UAV. That can be easily estimated through the fast Fourier transform (FFT), that is

$$Y_p[n] = \sum_{m=0}^{M_p - 1} y_p[n] e^{-j\frac{2\pi nm}{M_p}} \quad (4)$$

where $n = 0, 1, \ldots, M_p - 1$ with the number of FFT points $M_p$ in the $p$ th UAV under the same frequency resolution

$$\delta f_s = f_{s_p} / M_p \quad (5)$$

That means the bandwidth of buckets for each UAV is the same, which is the resolution for the proposed wideband spectrum acquisition method. Finally, each UAV distributes the result set $\mathbb{S}_p$ of the spectral peak search to others in the swarm including the index of the occupied bucket and its amplitude. Thus, the scheme has high efficiency and low information transmission capacity is presented.

*C. Coding and Decoding*

The sparse coding Fourier transform methods are radically different from the sparse Fourier transform. The difference lies in the way collision detection and resolution are handled. For the sparse Fourier transform, the idea is to use iterative between different sizes of bucketization. Once the sampling rate of UAVs has been selected, the frequency bucketization cannot be easily changed. The sparse coding Fourier transform uses the phase rotation property in the frequency domain, such that

$$f_{l_p} + k_p f_{s_p} = f_{l_p} \quad (6)$$

with the Nyquist zone index $k_p = \text{ceil}(f_s / f_{s_p})$, which follows the cycle with the period $f_{s_p}$. As such, $f_{l_p}$ is also the folding frequency of $f_{l_p} + k_p f_{s_p}$ sub-Nyquist sampled by $f_{s_p}$. To this end, the entire frequency range maps into

$$\mathbf{B}_{f_{l_p}} = \left[ \mathbf{b}_{f_{l_p}} \; \mathbf{b}_{f_{l_p}+f_{s_p}} \; \mathbf{b}_{f_{l_p}+2f_{s_p}} \ldots \mathbf{b}_{f_{l_p}+k_p f_{s_p}} \right]^T$$
$$= \left[ \mathbf{b}_{f_{l_p}}^0 \; \mathbf{b}_{f_{l_p}}^1 \; \mathbf{b}_{f_{l_p}}^2 \ldots \mathbf{b}_{f_{l_p}}^{k_p} \right]^T \quad (7)$$

where $(\cdot)^T$ is the transpose operator and $\mathbf{b}_{[\cdot]}$ denotes a vector of the folding spectrum for the different bands of $p$ th UAV. As a result, the spectrum is uniquely represented as

$$\mathbf{B} = \left( \mathbf{B}_{f_{l_1}} \; \mathbf{B}_{f_{l_2}} \; \mathbf{B}_{f_{l_3}} \cdots \mathbf{B}_{f_{l_p}} \right)^T \quad (8)$$

which is illustrated in Fig.3. The spectrum aliases equally spaced the digital frequencies through sub-sampling the input by $3\times$ and $4\times$ in the time domain. Specifically, it is clear that the digital spectrum maps input naturally into the code by cyclic continuation, as listed

$$\text{Codes} \begin{pmatrix} 3\times \\ 4\times \end{pmatrix} = \begin{pmatrix} 0 & 1 & 2 & 0 & 1 & 2 & 0 & 1 & 2 & 0 & 1 & 2 & | & 0 & \cdots \\ 0 & 1 & 2 & 3 & 0 & 1 & 2 & 3 & 0 & 1 & 2 & 3 & | & 0 & \cdots \end{pmatrix} \quad (9)$$

Frequencies $\;\;\;\boxed{0}\;\boxed{1}\;\boxed{2}\;\boxed{3}\;\boxed{4}\;\boxed{5}\;\boxed{6}\;\boxed{7}\;\boxed{8}\;\boxed{9}\;\boxed{10}\;\boxed{11}\;|\;\boxed{12}\;\cdots$

where the ambiguous frequencies are denoted with red color. Meanwhile, the maximum detectable dynamic range is the least common multiple for the set of the subsampling rate, which is the equivalent Nyquist sampling rates. Then, it is possible to resolve the frequencies unambiguously using the code

$$\mathbb{C} = \{ c_{p,q} | c_{p,q} = mod(q + \frac{M_p}{2}, M_p) - \frac{M_p}{2} \} \quad (10)$$

for the index of codes $q \in \{1, 2, \ldots, Q-1\}$, where the total number of codes $Q$ is as follows

$$Q = f_s / \delta f_s \quad (11)$$

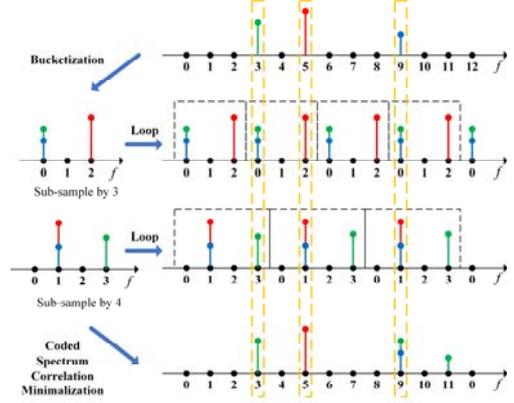

Fig. 3. Spectrogram of the NYFR output signal form the UAV swarm.

As we can see, each column of the coding sequence $\mathbb{C}$ makes up the sensing code for each channel $q$, which corresponds to the observed spectrum. Meanwhile, that is the index number of occupied buckets for each UAV $p$, which indicates the folded frequency for each sampler. Accordingly, a $Q$-dimensional binary matrix $\mathbf{c}$ is defined to distinguish the frequencies in the UAV, as

$$\langle \mathbf{c} \rangle_{c_{p,q}} = \begin{cases} 1, & c_{p,q} \in \mathbb{S}_p \\ 0, & else \end{cases} \quad (12)$$

where $\langle \cdot \rangle_{c_{p,q}}$ denotes the index of frequencies corresponding to the UAV, with the element being 1 for the occupied locations and 0 for the empty locations. Hence, the received information from the UAV swarm can be modeled as

$$\langle \mathbf{Y}_\mathbb{C} \rangle_{c_{p,q}} = \begin{cases} Y_p[c_{p,q}], & c_{p,q} \in \mathbb{S}_p \\ 0, & else \end{cases} \quad (13)$$

Obviously, the received signals from the UAV swarm can be related to the coding spectrum by

$$\mathbf{Y}_\mathbb{C} = \mathbf{B} \circ \mathbf{c} \quad (14)$$

where $\circ$ denotes the Hadamard product operator. Thus, the wideband spectrum can be acquired by calculating the spectrum correlation of each channel $q$ by its covariance matrix

$$\mathbf{R}_q = \mathbf{Y}_{\mathbb{C}_q} \times \mathbf{Y}_{\mathbb{C}_q}^H \quad (15)$$

where $(\cdot)^H$ is the conjugate transpose operator. In the $\mathbf{R}_q$, diagonal elements contain the self-power of each sampler output, while their cross-power is included in the rest of the elements. Knowing the elements of $\mathbf{R}_q$ is the power contained in the $q$ th channel between each UAV. All the UAVs have the same input component corresponding to the channel inside the code if the sensing code for the input is the index $q$. As such, every value of the spectrum correlation matrix contains the signal energy of interest. Naturally, the input component is present and different in only a portion of the paths for the other codes. Therefore, the element is implied as the noise power in the spectrum correlation matrix.

From the above discussion, the minimum element of the covariance matrix $\mathbf{R}_q$ is treated as the observed value of the signal power at the $q$ th channel. And then, the wideband spectrum can be sensed by ordering the minimum of the matrix $\mathbf{R}_q$ sequentially, as presented in Fig. 3. It is remarkable that the frequency with ambiguity comes into the same remainder of frequencies for each sampling path, which depends on the sampling rate. However, the multiple different aliasing guarantees that any two frequencies clashed in one sensing code will not collide in another, which is to the Chinese remainder theorem.

Furthermore, this is a rough estimation of the spectrum for the following signal processing. The sub-Nyquist sampling, as we know, will lose detection gain, which is mitigated by the channel gain and spatial gain through the filter of bucketing and beamforming. Then high-precision parameter estimation and real-time spectrum analysis are possible.

## IV. SIMULATION RESULTS

In this section, the simulation results are demonstrated to show the performance of the wideband spectrum acquisition for the UAV swarm. First, the UAV swarm with two nodes is implemented, each with a sampling rate of 3GHz and 4GHz, respectively. Thus, the equivalent Nyquist sampling rate for the spectrum acquisition system is 12GHz. To this end, assumed that there are $I$ signal components with identical powers randomly distributed in the frequency range of 0.35GHz to 12GHz. The capture time of inputs sets to 1ms. To evaluate performance, the state-of-the-art generalized coprime sampling method [17] is compared to the proposed method under the same conditions using the relative root mean square error (Relative RMSE), defined as

$$\text{Relative RMSE} = \frac{1}{f_s}\sqrt{\frac{1}{KI}\sum_{k=1}^{K}\sum_{i=1}^{I}(\hat{f}_{c_i}(k) - f_{c_i})^2} \quad (16)$$

where $\hat{f}_{c_i}(k)$ is the estimated frequency of $f_{c_i}$ for the $k$ th Monte Carlo trial and $K$ is empirically set to 500.

Wide an input SNR of 0dB, the spectrum is acquired with different bucketization bandwidths of 100MHz, 10MHz, and 1MHz. Meanwhile, there are $I=4$ signals at frequencies of 0.95GHz, 7.56GHz, 3.37GHz, and 10.5GHz, whose modulation types are monopulse, binary phase shift keying, and linear frequency modulation with bandwidths of 100MHz and 1GHz, respectively. As can be seen, the performance of the spectrum acquisition method and the noise level improves as the bandwidth of frequency bucketization decreases. Hence, the smaller frequency resolution is achieved at the cost of increased data transmission and computational intensity. Hence, there is a trade-off between system resources and performance.

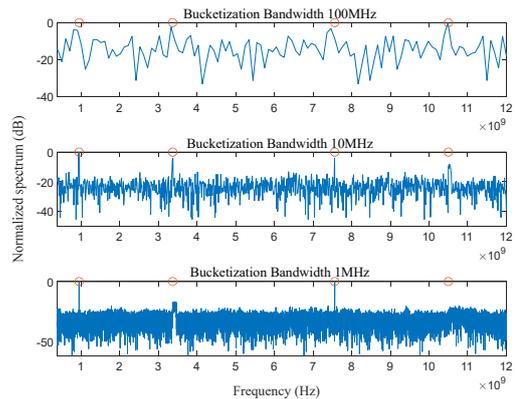

Fig. 4. Spectrum acquisition with different bucketization bandwidth for UAV swarm ($I=4$ and $\text{SNR}=0\text{dB}$).

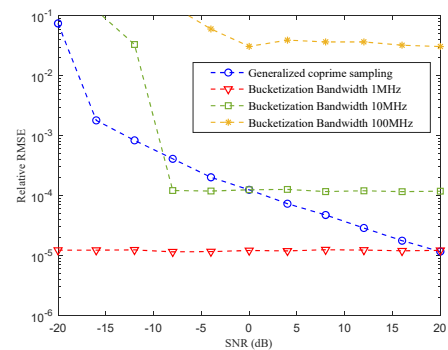

Fig. 5. Relative RMSE versus SNR ($I=1$).

The Relative RMSE results comparing the proposed method with the generalized coprime sampling scheme are shown in the Fig. 5, where $I=1$ is assumed with different SNRs. The results display that the RMSE tends towards stability after reaching a certain SNR for different bucketization bandwidths, which is different from the generalized coprime sampling scheme. Note that, the method is able to acquire the spectrum without placing any sparsity constraints on the monitoring spectrum. However, the aliasing noise increases with increasing spectrum density, which impacts the system's dynamic range. Therefore, spatial gain from the number of nodes in the UAV swarm, the channel gain from the bandwidth of the frequency bucketization, and the monitoring time are utilized to mitigate the loss of gain from sub-sampling.

## V. CONCLUSION

A wideband spectrum acquisition technique based on sparse coding Fourier transform has been developed for the UAV swarm scenario. By introducing the frequency bucketization and estimation from the sparse Fourier transform theory, each UAV can perform subsampling sensing of the entire spectrum. The proposed coding and encoding scheme successfully solves collision and iteration without sparsity constraints. Moreover, the UAV swarm can be directly reconfigured even if one of the nodes is lost, at the cost of losing the monitoring bandwidth. In the future, the studies will focus on the theoretical analysis of boundary conditions for sparse coding Fourier transform, the perturbation effect on the system, and the selection of the parameters for the UAV swarm.